\documentclass[manuscript=article,layout=twocolumn]{achemso}

\usepackage{amsmath}
\usepackage{xcolor}
\usepackage{braket}
\usepackage{array}
\usepackage{graphicx}
\usepackage{subfig}
\usepackage{caption}
\usepackage[singlelinecheck=false]{caption}
\usepackage[normalem]{ulem}
\usepackage{comment}
\graphicspath{ {./figs/} }

\title{Fast and Accurate Excitation Energies from Density Matrix Renormalization Group Calculations Improved by Machine Learning}
\author{Pavlo Golub}
\email{pavlo.golub@jh-inst.cas.cz}
\author{Libor Veis}
\email{libor.veis@jh-inst.cas.cz}
\affiliation[jh-inst]
{Department of Theoretical Chemistry, Heyrovsk\'y Institute of the Czech Academy of Sciences, Prague 18223, Czech Republic.}

\date{02.2026}

\begin{document}

\maketitle

\begin{abstract}
 An estimation of optoelectronic properties is a crucial necessity and an ongoing challenge in modern functional material science. This involves accurate accessing of ground and excited states in systems with complex electronic structure that are often too computationally expensive for accurate many-body methods. Building on our previous work on ground states [J. Phys. Chem. Lett. 2025, 16, 3295-3301], we present an efficient and cost-effective approach for evaluating electronic excitations in functional materials, combining the density matrix renormalization group method as a complete active space solver with machine learning techniques. We demonstrate its performance on $\pi$-electron-correlated systems, namely polycyclic aromatic compounds. The transferability and effectiveness of the derived machine learning model are demonstrated on a number of challenging examples with up to 34 $\pi$-electrons.
\end{abstract}

\section{Introduction}

Precise information on electron excitation profiles of functional molecules is crucial for understanding the underlying photochemical processes and for designing and exploring materials with desired optical properties. In this context, particular attention has been devoted to metal-free organic systems in which desirable chemical processes, such as catalysis, charge-transfer, and light-emitting, are triggered by  photon absorption \cite{Romero-2016,Liu-Li-2020,Zhang-Wang-2022,Cabanero-Rovis-2025}. As such systems often exhibit an intricate electronic structure, accurate treatment of electron correlation becomes essential for a reliable description of their electronic transitions. 

Not surprisingly, machine learning (ML) methods have been widely applied to mitigate the complexity of the task and consequently to allow a fast screening of the chemical compound space. The streamline of such ML approaches is to avoid in the process of prediction the use of expensive electronic structure methods. With this aim a lot of efforts have been devoted to establishing a correspondence between molecular geometrical/structural parameters and their quantum properties. An atomic environment is represented with a number of descriptors derived from system internal coordinates, chemical composition, presence of special constituent elements such as double/triple bonds, electron conjugated sub-parts, etc\cite{Rupp-Tkatchenko-2012,Hansen-Biegler-2015,Faber-Christensen-2018,Huang-Lilienfeld-2020,Rogers-Hahn-2010,Behler-2011,Bartok-Payne-2010,Bartok-Kondor-2017,Willat-Musil-2019,Gastegger-Schwiedrzik-2018,Drautz-2019,Jaeger-Fulle-2018,Guha-Velegol-2023,Zalte-Pang-2025}. Sometimes in the learning model such structural descriptors are used with an admixture of quantum molecular parameters (so-called quantum-chemical descriptors) typically obtained at the level of density functional theory (DFT): oscillator strength, dipole momentum, information on frontier molecular orbitals, etc\cite{Yap-2011,Wang-Jin-2012,Mao-Yao-2024}.

% Overall, side by side, molecular descriptors are routinely used in combination with different regression and generative models in efforts to find compounds with desirable photo properties (absorption/emission spectra, quantum yield)\cite{Gupta-Chakraborty-2021,Ye-Huang-2020,Tan-Li-2023,Mahato-Kumar-2024,Bi-Jiang-2024,Peng-Shen-2025}, and new multiproperty prediction models, tools, and databases are developed\cite{Yang-Swanson-2019,Ju-Bai-2021,Chen-Bononi-2022,Shao-Liu-2022,Hung-Ye-2023,Jung-Jung-2024,Song-Xiong-2025,Souza-Duarte-2025,Nguyen-Le-2025,Zhao-Li-2025,Wang-Wu-2025}.

Nevertheless, there are serious challenges to this approach. As the structural space is infinite, the number of possible geometrical variations is countless, severely diminishing the possibility of capturing all important structure-property dependencies. The databases include tens to hundreds of thousands of samples, molecular/atomic/structural descriptors form large multidimensional feature spaces, and there is still the problem of extrapolating model performance beyond the systems that are structurally similar to test/training samples\cite{Gupta-Chakraborty-2021,Chen-Bononi-2022}.

In addition, since organic systems of interest are often characterized by strongly correlated patterns arising mainly from the distribution of $\pi$-electrons, the accuracy of the structure-property mapping might be questionable. Time-dependent density functional theory (TDDFT)\cite{Runge-Gross-1984} is widely used as a reference method as it offers an acceptable tradeoff between quality and cost of calculations for thousands of samples. However, it heavily depends on the applied functional, often tends to overestimate transition energies, and experiences apparent accuracy problems in charge-transfer states\cite{Peach-Benfield-2008,Laurent-Jacquemin-2013,Kang-Kim-2024}. Different variants of equation of motion and linear response coupled clusters with singles and doubles (EOM-CCSD, CCSD-LR)\cite{Sekino-Bartlett-1984,Koch-Jorgensen-1990,Nooijen-Bartlett-1997,Berraud-Pache-Neese-2020} outperform TDDFT, however, they are still lacking systematicity, as the error can jump from 0.0 up to and above 0.2 eV with reference to highly accurate electronic structure methods or experiment\cite{Kang-Kim-2024,Kozma-Tajti-2020}. There is no guaranty that the potentially promising candidates selected by the model will actually have predicted photochemical properties and that important molecules will not be overlooked. 

Therefore, the use of advanced electronic structure methods for strong correlation appears to be an important component of ML models. However, given their cost, a way to recover their high-level accuracy from low-level and cheaper calculations should be found, which brings one to the concept of $\Delta$-ML\cite{Ramakrishnan-Dral-2015}. In $\Delta$-ML, a model is trained to predict the difference ($\Delta$) between a target property obtained with a high-level method and its counterpart obtained with a computationally cheaper, low-level method. The predicted correction is then added to the low-level result, providing an approximation to the high-level result at a substantially reduced computational cost. A successful $\Delta$-ML model could serve as a reliable reference to the aforementioned models based on molecular descriptors, or could be applied as an independent predictor utilizing electronic structure information from low-level input. 

Until now, there have been comparatively few studies in $\Delta$-learning on excited states in organic photochemistry. 22,000 small organic molecules have been subject to refinement of the TDDFT level to the second-order approximate coupled-cluster (CC2) theory\cite{Ramakrishnan-Hartmann-2015}. Relying on Coulomb-matrix molecular representations\cite{Rupp-Tkatchenko-2012,Hansen-Biegler-2015} as learning inputs, the model predicted $S_0 \rightarrow S_1$ transition energies and the corresponding oscillator strengths, in general, improving DFT, but not always to a satisfactory level, possibly because of contradictions between CC2 and DFT, as both methods could yield a different order of states. Only an insignificant improvement in prediction of oscillator strengths has been detected with increasing training set size\cite{Ramakrishnan-Hartmann-2015}. There are also attempts to learn the TDDFT quality output from semiempirical\cite{Coxson-Omar-2026} or less accurate DFT\cite{Verma-Rivera-2022} approaches with mean absolute error mainly less than 0.5 eV. This, however, brings again the discussion regarding the performance of TDDFT, although the possibility to extend the learning protocol to more accurate correlated methods has been shown\cite{Verma-Rivera-2022}. To the same category of ML might be classified the protocols for using molecular integrals from low-level theories as leaning components\cite{Welborn-Cheng-2019,Chen-Yam-2023}.

A potential use of the density matrix renormalization group (DMRG) \cite{White-1992,White-1993,Schollwock-2005, chan_review, yanai_review, Szalay2015, reiher_perspective}, theory in the context of $\Delta$-ML is a promising alternative. Indeed, in this sense DMRG looks appealing, as it provides systematically improvable approximations to low-energy eigenstates of a full configuration interaction (FCI) solution for a wave function in a chosen complete active space (CAS). The accuracy of the approximation can be chosen arbitrarily and controlled by the truncation error (TRE) (will be discussed in section Theory). Recently, we have shown that the use of state-specific TRE and quantum information-based measures (orbital entropies and mutual information) allows for highly accurate predictions of ground-state energies within a given spin sector, including singlet and triplet states, for out-of-model examples with localized $\pi$-active spaces \cite{Golub-2025}. Although the approach requires much more insight from the user as it depends on proper CAS selection and post-DMRG wave function analysis, it benefits from the possibility to make predictions in the FCI limit and very moderate requirements on the number of model samples and input parameters.

The aim of this work is to significantly extend the aforementioned development to excited states, specifically to same-spin excitations within general $\pi$-active spaces, thus targeting singlet–singlet and triplet–triplet transitions. 
%Also an effect of dynamical correlation correction on-top of both the predicted and approaching FCI limit solutions will be investigated. \lv{Is it true?} 

In what follows, we briefly review the theoretical aspects of DMRG relevant to our developments, describe the machine learning model, and assess its performance on selected polycyclic aromatic compounds.

\section{Theory}
\subsection{Density Matrix Renormalization Group}

Starting from the FCI wave function expansion in a given CAS, expressed in the occupation basis representation

\begin{equation}
  | \Psi \rangle = \sum_{\boldsymbol{\sigma}}^{n} c_{\boldsymbol{\sigma}} | \boldsymbol{\sigma} \rangle,
\end{equation}

\noindent
where $ | \boldsymbol{\sigma} \rangle = | \sigma_1 \sigma_2 \ldots \sigma_n \rangle$ represents an occupation-number basis state of the active-space orbitals, with each orbital having four possible occupation states $| 0 \rangle, | \downarrow \rangle, | \uparrow \rangle, | \downarrow \uparrow \rangle$, DMRG in its matrix product state (MPS) formulation seeks a factorization of the coefficient tensor $c_{\boldsymbol{\sigma}}$\cite{Schollwock-2011,Schollwock-2011-2}.

Briefly, the procedure starts by reshaping the coefficient tensor into a $(4 \times 4^{n-1})$ matrix, $c_{\boldsymbol{\sigma}} = \overline{c}_{\sigma_1;\sigma_2 \cdots \sigma_n}$, to which singular value decomposition (SVD) is then applied.

\begin{equation}
    \label{SVD1}
    c_{\boldsymbol{\sigma}} = \overline{c}_{\sigma_1;\sigma_2 \cdots \sigma_n} = \sum_{k_{1}} U_{\sigma_1; k_{1}} S_{k_{1};k_{1}} V_{k_{1};\sigma_2 \cdots \sigma_n}
\end{equation}

\noindent
This yields a product of matrices with dimensions $(4 \times k_1)$, $(k_1\times k_1)$, and $(k_1 \times 4^{n-1})$, respectively. For an SVD of an $(m \times n)$ matrix, the number of singular values is $\min(m,n)$. Therefore, in the present case, $k_1 = \min(4,4^{n-1})$.

Setting
$A_{k_1}^{\sigma_1}
= U_{\sigma_1;k_1} S_{k_1;k_1}$,
the remaining tensor
$V_{k_1;\sigma_2 \cdots \sigma_n}$
is reshaped into a $(4k_1 \times 4^{n-2})$ matrix by combining the
indices $k_1$ and $\sigma_2$. A second SVD is then performed,

\begin{equation}
    \label{SVD2}
    c_{\boldsymbol{\sigma}}
    =
    \sum_{k_1}
    A_{k_1}^{\sigma_1}
    \left[
    \sum_{k_2}
    U_{k_1 \sigma_2;k_2}
    S_{k_2;k_2}
    V_{k_2;\sigma_3 \cdots \sigma_n}
    \right].
\end{equation}

\noindent
The three factors in the second SVD have dimensions
$(4k_1 \times k_2)$, $(k_2 \times k_2)$, and
$(k_2 \times 4^{n-2})$, respectively, where
$k_2 = \min(4k_1,4^{n-2})$.
Absorbing the singular values into the left factor, analogously to
the first SVD, we define
\begin{equation}
    A_{k_1 k_2}^{\sigma_2}
    =
    U_{k_1 \sigma_2;k_2} S_{k_2;k_2},
\end{equation}
which gives
\begin{equation}
    c_{\boldsymbol{\sigma}}
    =
    \sum_{k_1 k_2}
    A_{k_1}^{\sigma_1}
    A_{k_1 k_2}^{\sigma_2}
    V_{k_2;\sigma_3 \cdots \sigma_n}.
\end{equation}

The procedure is repeated until the last, $n$th, orbital is reached. The resulting MPS representation of the wave function takes the form

\begin{equation}
  \label{mps_factorization}
  | \Psi \rangle = \sum_{\boldsymbol{\sigma}} \sum_{k_1 \ldots k_{n-1}} A_{k_1}^{\sigma_1} A_{k_1 k_2}^{\sigma_2} A_{k_2 k_3}^{\sigma_3} \cdots A_{k_{n-1}}^{\sigma_n} | \boldsymbol{\sigma} \rangle.
\end{equation}

\noindent
The above factorization corresponds to a right-canonical construction,
as the singular-value matrices are successively absorbed into the
left factors. The particular canonical form is not unique and can be
changed by gauge transformations without altering the represented
wave function.

Each matrix $A^{\sigma_j}$, with the boundary matrices $A^{\sigma_1}$ and $A^{\sigma_n}$ reducing to vectors,
corresponds to a particular orbital $j$.
As follows from the factorization described above, the bond dimensions are
$k_j = \min(4^j,4^{n-j})$ and therefore increase exponentially toward
the middle of the chain before decreasing symmetrically thereafter.
Retaining the full bond dimensions would therefore result in an exact
MPS representation of the FCI wave function but would preserve its
prohibitive exponential complexity. The key feature of DMRG is that
these dimensions can be truncated at a chosen maximum value, resulting
in an approximate solution of the FCI problem. The accuracy of this
approximation is controlled by the maximum retained bond dimension,
commonly referred to as the bond dimension, $D$.

The practical two-site DMRG algorithm\cite{Wouters-Neck-2014} takes a pair of tensors and considers all tensors to the left of them as a single (left, \textit{L}) block and all tensors to the right as another single (right, \textit{R}) block:
\begin{equation}
  \label{tensor_split}
  \begin{split}
 \overset{L \text{ block}}{ \boxed{ A_{k_1}^{\sigma_1} \cdots A_{k_{l-2} ~ k_{l-1}}^{\sigma_{l-1}} }  } ~\cdot~
 A_{k_{l-1} ~ k_{l}}^{\sigma_l} A_{k_{l} ~ k_{l+1}}^{\sigma_{l+1}} ~ \cdot
 \\ \overset{R\text{ block}}{\boxed{ A_{k_{l+1} ~ k_{l+2}}^{\sigma_{l+2}} \cdots A_{k_{n-1}}^{\sigma_n} } }.
   \end{split}
\end{equation}

The optimization procedure is organized into sweeps, each consisting
of a sequence of local optimization steps. At each step, two neighboring
MPS tensors (which together can be seen as two-site optimization window) are contracted to form a four-index two-site tensor,
\begin{equation}
  \label{Fused_matrix}
  B_{k_{l-1}k_{l+1}}^{\sigma_l\sigma_{l+1}}
  =
  \sum_{k_l}
  A_{k_{l-1}k_l}^{\sigma_l}
  A_{k_lk_{l+1}}^{\sigma_{l+1}},
\end{equation}
\noindent
which is subsequently optimized while the tensors associated with the
$L$ and $R$ blocks are kept fixed.

The optimization of the two-site tensor $B$ is formulated as a
projected Schr\"odinger equation in the effective basis
\begin{equation}
  |L_{k_{l-1}}\rangle
  |\sigma_l\rangle
  |\sigma_{l+1}\rangle
  |R_{k_{l+1}}\rangle,
\end{equation}
\noindent
formed by the renormalized states of the $L$ and $R$ blocks and the
local occupation states of the two optimized sites. Projection of the
Hamiltonian onto this basis yields the effective eigenvalue problem
\begin{equation}
  \label{effective_eigenvalue}
  \sum_{\substack{
  k'_{l-1},k'_{l+1}\\
  \sigma'_l,\sigma'_{l+1}
  }}
  \left(H_{\mathrm{eff}}\right)^{
  \sigma_l\sigma_{l+1},\,
  \sigma'_l\sigma'_{l+1}
  }_{
  k_{l-1}k_{l+1},\,
  k'_{l-1}k'_{l+1}
  }
  B_{k'_{l-1}k'_{l+1}}^{
  \sigma'_l\sigma'_{l+1}
  }
  =
  E
  B_{k_{l-1}k_{l+1}}^{
  \sigma_l\sigma_{l+1}
  },
\end{equation}
\noindent
where $H_{\mathrm{eff}}$ is the Hamiltonian projected onto the
renormalized two-site basis. Solving Eq.~\ref{effective_eigenvalue}
yields the optimized two-site tensor $B$, corresponding to the
targeted eigenstate (or few eigenstates).

After solving the effective eigenvalue problem, the optimized two-site
tensor $B$ is reshaped into a matrix and factorized by SVD:
\begin{equation}
  \label{SVD_on_matrix}
  B_{k_{l-1}k_{l+1}}^{\sigma_l\sigma_{l+1}}
  =
  \sum_{k_l}
  U_{k_{l-1}k_l}^{\sigma_l}
  S_{k_l}
  V_{k_lk_{l+1}}^{\sigma_{l+1}}.
\end{equation}

\noindent
During the DMRG optimization, the canonical form is adapted to the
direction of the sweep, such that the orthogonality center follows the
two-site optimization window. During a forward sweep from left to right,
$U_{k_{l-1}k_l}^{\sigma_l}$ is assigned to the left tensor of the
optimized pair, $A_{k_{l-1}k_l}^{\sigma_l}$, while the product
$S_{k_l}V_{k_lk_{l+1}}^{\sigma_{l+1}}$ is assigned to the right
tensor, $A_{k_lk_{l+1}}^{\sigma_{l+1}}$.
The $L$ block is then enlarged
by incorporating the optimized tensor at site $l$, whereas the $R$
block is correspondingly reduced. The optimization window is shifted
by one site, resulting in the next two-site pair
\begin{equation}
  A_{k_lk_{l+1}}^{\sigma_{l+1}}
  A_{k_{l+1}k_{l+2}}^{\sigma_{l+2}},
\end{equation}
\noindent
to which the same optimization procedure is applied.

A complete sweep proceeds through the chain from left to right and then reverses direction at the end of the chain. Forward and backward sweeps are repeated until the chosen convergence criterion is satisfied.

At each local optimization step, the SVD of the optimized two-site tensor
provides a natural way to truncate the MPS bond dimension. According to
the Eckart--Young theorem, the best rank-$D$ approximation of a matrix
in the least-squares sense is obtained by retaining only its $D$ largest
singular values. Thus, after the decomposition in
Eq.~\ref{SVD_on_matrix}, only the $D$ largest singular values $s_i$ and
the corresponding columns of $U$ and rows of $V$ are retained.

For a normalized MPS in canonical form, the singular values satisfy
\begin{equation}
    \sum_i s_i^2 = 1.
\end{equation}
The error introduced by truncating the bond dimension to $D$ is therefore
quantified by the discarded weight,
\begin{equation}
    \label{TRE}
    \mathrm{TRE}
    =
    \sum_{i>D} s_i^2
    =
    1 - \sum_{i=1}^{D} s_i^2.
\end{equation}
This quantity defines the truncation error (TRE). Equivalently, the
squared singular values $s_i^2$ are the eigenvalues of the reduced
density matrix associated with the corresponding bipartition of the
MPS. Thus, truncation based on the SVD is equivalent to retaining the
$D$ most significant eigenstates of the reduced density matrix
\cite{Schollwock-2005}.

TRE is particularly useful in the present approach, as it can be
employed in extrapolation schemes to recover accurate results from
less precise but computationally less expensive calculations
\cite{Olivares-Amaya-Hu-2015,Golub-2025}.

\subsection{Machine Learning Model}

\begin{figure}[!t]
  \includegraphics[width=8.5cm]{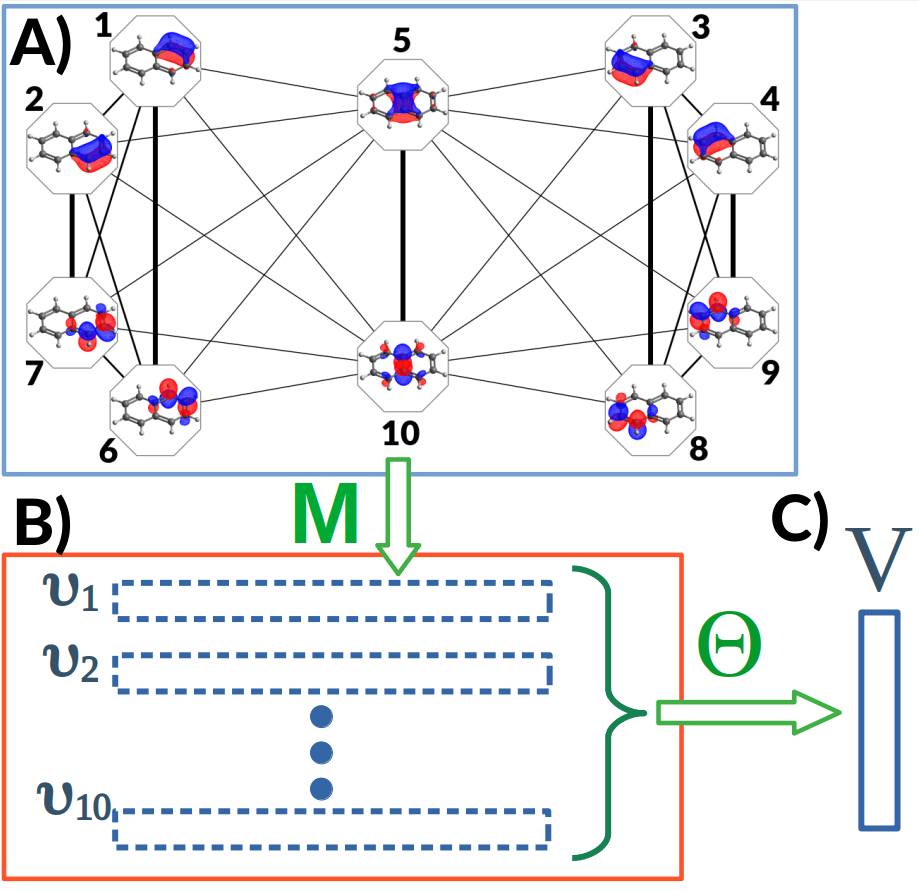}
  \caption{The scheme of MPGNN on example of naphthalene. A) The $\pi$-active space (10e,10o) is represented as a graph. Each orbital is a node. Nodes are connected according to the value of  mutual information $I_{ij}$ (Eq. \ref{mut_info}) between orbitals. In this example the threshold value of $I_{ij}$ is 0.02. B) For each node a vector of feature values $\upsilon$ is formed according to the message representation (Eq. \ref{message_formation}). C) The final feature representation of the molecule, vector $V$, is formed by applying aggregation function $\Theta$ on the vector node representations $\upsilon$.}
  \label{gnn_scheme}
\end{figure}

The Message Passing Graph Neural Network (MPGNN) model for $\pi$-active spaces, which we have recently presented \cite{Golub-2025}, treats each orbital in a given space as a node, while the single-orbital entanglement entropy, $s^{(1)}$, is treated as a node-specific feature. The single-orbital entropy measures the entanglement between a single orbital and the remaining orbital space. It is defined by the general $N$-orbital von Neuman entropy \cite{Legeza-Solyom-2003,Legeza-Solyom-2004,Rissler-Noack-2006}:

\begin{equation} \label{N_entrop}
    s^{(N)} = -\sum_{\sigma}{w_{\sigma;1...N} ~ \textrm{ln} (w_{\sigma;1...N})},
\end{equation}

\noindent
where $w_{\sigma;1,...N}$ represent the eigenvalues of the reduced $N$-orbital density matrix.
The presence of an edge between two nodes is determined by a threshold value of mutual information:

\begin{equation} \label{mut_info}
    I_{ij} = s^{(1)}_{i} + s^{(1)}_{j} - s^{(2)}_{ij}.
\end{equation}

\noindent
where $s^{(2)}_{ij}$ is the two-orbital entropy that measures the entanglement between a pair of orbitals $\{ij\}$ and the remaining orbital space. Mutual information therefore measures the correlation between two specific orbitals. If an edge exists (i.e. $I_{ij}$ exceeds a predefined threshold), the corresponding mutual information is considered as an edge-specific feature (Fig. \ref{gnn_scheme},A). 

 The feature information on the connected nodes, together with the corresponding feature information on the edges, is used to build a feature vector representation of each node that accounts for its environment:

\begin{equation} \label{message_formation}
    \upsilon_{i}^{m+1} = M(\upsilon_{i}^{m}, \upsilon_{j}^{m}, e_{ij}).
\end{equation}

\noindent
Here $M$ defines the process of message formation, which involves the aggregation of node ($\upsilon$) and edge ($e$) features of all nodes $j$ connected to $i$, and the concatenation of the output vector representations $\upsilon_{i}^{m}$, $ \upsilon_{j}^{m}$ and $ e_{ij}$ at step $m$ to form a new representation (message) $ \upsilon_{i}^{m+1}$ (Fig. \ref{gnn_scheme},B). Generally this message is updated by a function ($F_{upd}$) such as the multi-layer perceptron to form a new state of the node $i$. However, it has been shown\cite{Golub-2025}  that, regarding the type of considered systems, $F_{upd}$ can be a simple identity function without loss of learning accuracy.

Combining vector representations $\upsilon_{i}$ of all nodes using an aggregation function $\Theta$ (such as the mean function) provides a vector representation of the system, $V$ (Fig. \ref{gnn_scheme},C). Further, this final representation, augmented with system specific features such as TRE, can be used for learning system properties.

\begin{figure}[!t]
  \includegraphics[width=8cm]{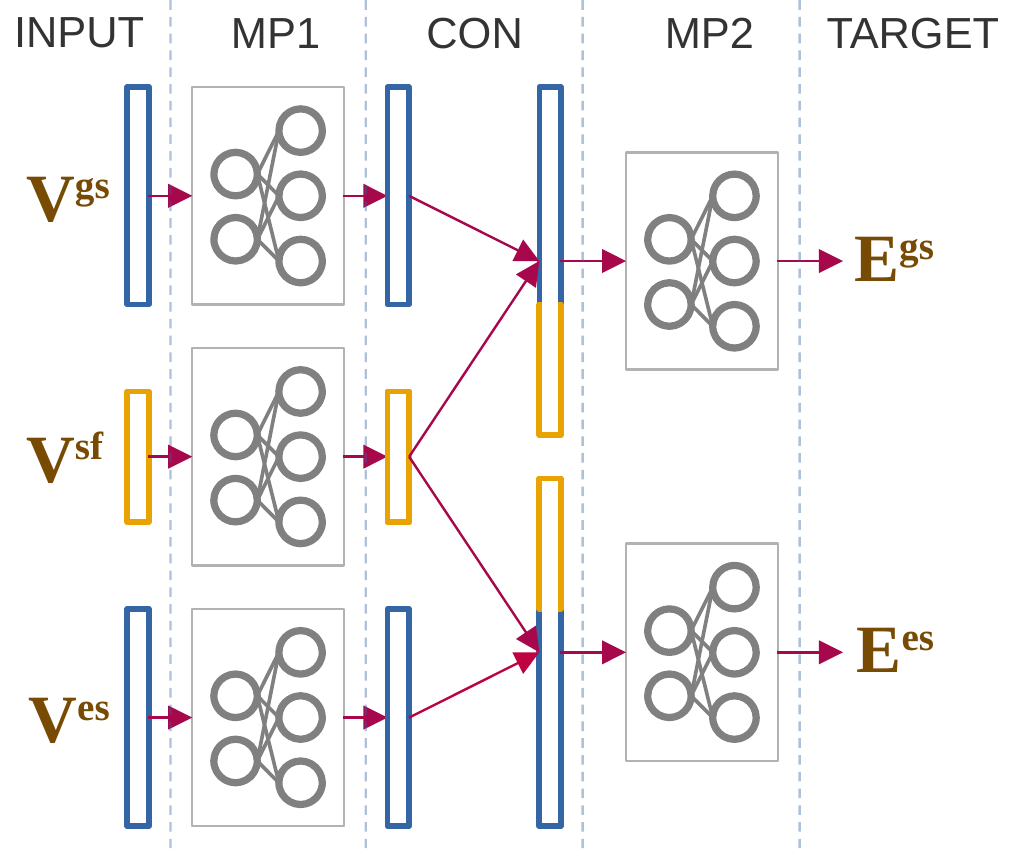}
  \caption{The scheme of learning workflow. Input feature vectors $V^{gs}$, $V^{es}$, and $V^{sf}$ are passing through multiperceptron networks (stage \textit{MP1}). Output vector $V^{sf}$ is concatenated to the output vectors $V^{gs}$ and $V^{es}$ (stage \textit{CON}). Concatenated vectors are passing through another multiperceptron networks (stage \textit{MP2}) to learn against target energies $E^{gs}$ and $E^{es}$.}
  \label{learning_scheme}
\end{figure}

Our current model deals with two states per molecule simultaneously, the ground state and the excited state; therefore, for each state, a separate vector representation is built, $V^{gs}$ and $V^{es}$ respectively. The learning is performed on two target energies, $E^{gs}$ and $E^{es}$, which results in basically two parallel learning paths for the ground state and the excited state per molecule (see the scheme in Fig. \ref{learning_scheme}). In the process of learning, both vectors $V^{gs}$ and $V^{es}$ are augmented with a vector of system-dependent features, $V^{sf}$. The last is passing through an additional multi-layer perceptron unit prior to the augmentation. In the process of backpropagation, the weights of this unit are updated depending on the gradients from both learning paths, thus allowing them to indirectly affect each other.

\begin{figure}[!t]
  \includegraphics[width=8cm]{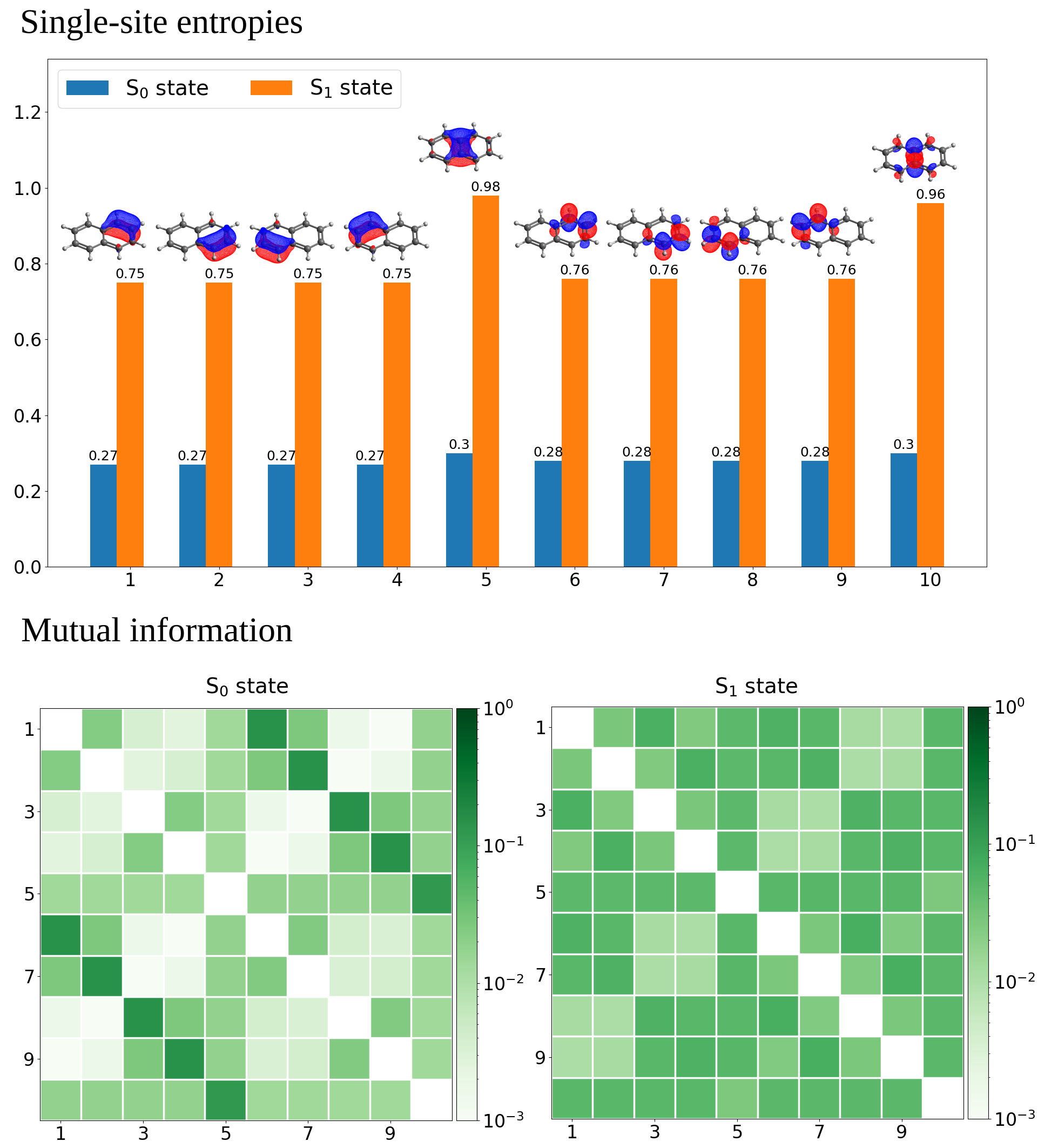}
  \caption{The difference of singe-site entropy and mutual information between the ground and excited states; an example of naphthalene in (10e, 10o) $\pi$-valence active space.}
  \label{S1_S2_diff}
\end{figure}

In the present model, vector $V^{sf}$ includes state averaged TRE, taken as a mean value over state averaged TREs in the middle of the chain for the last sweep in the convergence procedure, and a number of metrics for graph comparison, applied to initial graph representations of ground and excited states. In our case, the problem of general graph comparison reduces to comparing graph structures with complete node correspondence, differing only in number of edges and the magnitude of features on nodes and edges. Indeed, the difference in the informational measure between the ground and excited states is generally manifested in higher values for $s^{(1)}$ and $I$ for the excited state (Fig. \ref{S1_S2_diff}). We found several relevant measures: 

\begin{itemize}
    \item The mean degree of nodes, that represents the average number of edges per node.
    \item The Szymkiewicz-Simpson coefficient\cite{Szymkiewicz-1934} that, applied to our specific task, is the number of common edges between two graphs divided by the total number of edges in the smaller graph (generally the one that corresponds to the ground state).
    \item Frobenius distance between the corresponding adjacency matrices.
\end{itemize}

Those measures provide a noticeable improvement to the learning procedure by lowering the mean error on training and validation sets simultaneously.

\section{Results and discussion}

\subsection{Test dataset}

\begin{figure}[!t]
  \includegraphics[width=8cm]{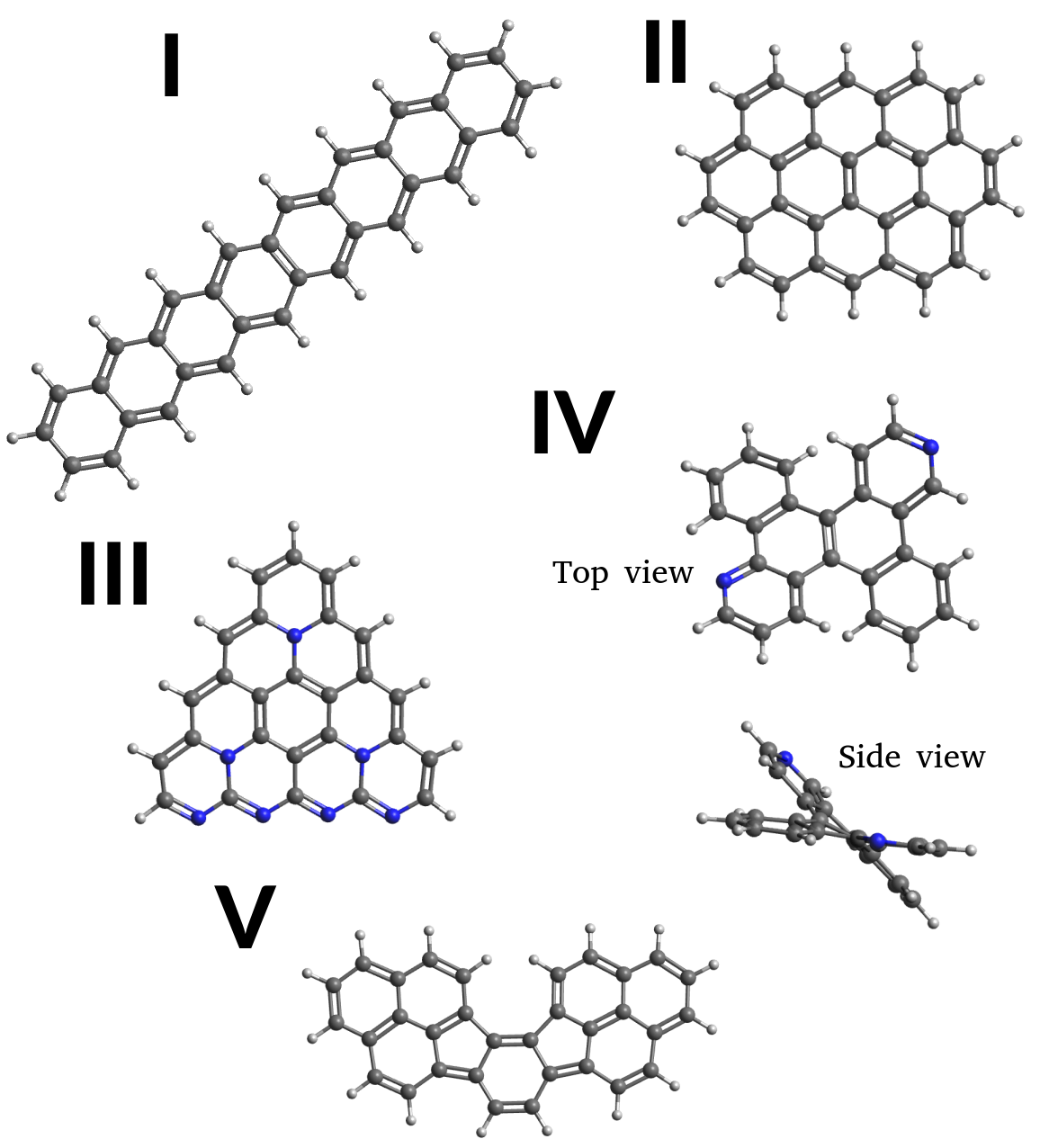}
  \caption{Test set of polycyclic hydrocarbons: {\bf{I}} -- C$_{34}$H$_{20}$, octacene; {\bf{II}} -- C$_{32}$H$_{14}$, ovalene; {\bf{III}} -- C$_{26}$N$_{7}$H$_{11}$, aza-analogue of [4]triangulene; {\bf{IV}} -- C$_{24}$N$_{2}$H$_{14}$, 4,11-diazadibenzo[\textit{g,p}]chrysene; {\bf{V}} -- C$_{32}$H$_{16}$.}
  \label{test_compounds}
\end{figure}

The test dataset has been selected with general motivation to evaluate transferability of the model on different types of hydrocarbons. Having in common $\pi$-electron states as driving force for strongly correlation effects, they significantly differ in molecular topology and underlying electronic structures. We mainly emphasize on peri-fused polycyclic aromatic hydrocarbons (PAHs), where the constituent rings are fused through more than one face, in contract to laterally fused PAHs presented in the train dataset (Fig. \ref{test_compounds}). 

Octacene (Fig. \ref{test_compounds}, I) is the only non peri-fused example in the test dataset. It is structurally similar to the examples in the train dataset but is extended compared to them. It belongs to the family of oligoacenes and is predicted to have a singlet ground state with biradical character\cite{Bendikov-Duong-2004,Hachmann-Dorando-2007}.

Ovalene (Fig. \ref{test_compounds}, II) is a flat compound of ten peri-fused benzene rings featuring six zigzag edges and $D_{2h}$ symmetry. It is predicted to posses global aromaticity, as calculated ring current of participating outer $\pi$-electrons is diatropic. Still the aromaticity of the inner part is predicted to be lower since the induced ring current of the inner part is weaker\cite{Chen-Schollmeyer-2019}. Recent research suggests that the $S_1 \rightarrow S_0$ transition in ovalene is accidentally forbidden due to destructive interference of the corresponding transition dipole moments\cite{Wega-Vauthey-2026}.

Aza-analogue of [4]triangulene (Fig. \ref{test_compounds}, III) serves for the evaluation of the model performance on hetero-PAHs. Pure hydrocarbon triangulenes are open-shell radicals that have higher and higher ground state spin numbers with increasing size: thus, [3]triangulene is characterized by a triplet ($S=1$) ground state, while [4]triangulene has a quartet ($S=3/2$) ground state\cite{Zeng-Wu-2021}. However, substitution of carbon atoms with nitrogen atoms (or other heteroatoms such as boron) changes the electronic structure of hydrocarbons and stabilizes in lower-spin or even closed-shell structures\cite{Bai-Zhang-2026}: for example,  substitution of the central carbon atom in [3]triangulene by a nitrogen changes the spin multiplicity in the ground state, making it doublet according to theoretical calculations \cite{Sandoval-Salinas-2019,Wang-Berdonces-Layunta-2022}.  

4,11-Diazadibenzo[\textit{g,p}]chrysene (Fig. \ref{test_compounds}, IV) is another example of hetero-PAH. It features a chrysenic core fused with two additional aromatic rings, where nitrogen atoms replace carbons at the 4 and 11 positions. Unlike previous examples and most of the model dataset members, it is an example of non-planar $\pi$-conjugated system, where non-planarity is caused by the repulsion of hydrogen atoms. Generally, twisted geometry weakens $\pi$-bonds, since $\pi$-orbitals are no longer orthogonal to the molecular plane and hence to $\sigma$-orbitals, leading to enhancement of $s$-character in $\pi$-bonding\cite{Haddon-1988,Kumar-Aggarwal-2020}. This weakening is reflected by the single-orbital entropy values, since the $s^{(1)}$value for the $\pi$-orbital localized on two central atoms is approximately in $2.5$ times smaller than the $s^{(1)}$ values for all other $\pi$- and $\pi^{*}$-orbitals.

The bis(phenalenyl) system linked by benzene (Fig. \ref{test_compounds}, V) is the last test example. It is seen as two phenalenyl radicals with delocalized unpaired electron partially stabilized by $\pi$-system, serving in the same time as connector.\cite{Sun-Wu-2012} Theoretical considerations based on natural orbital analysis suggest its significant biradical character.\cite{Kubo-Sakamoto-2005} Similar to the previous system, it has non-planar geometry according to B3LYP DFT geometry optimization with cc-pVTZ basis, although much less pronounced. 

\begin{figure}[!t]
  \includegraphics[width=8cm]{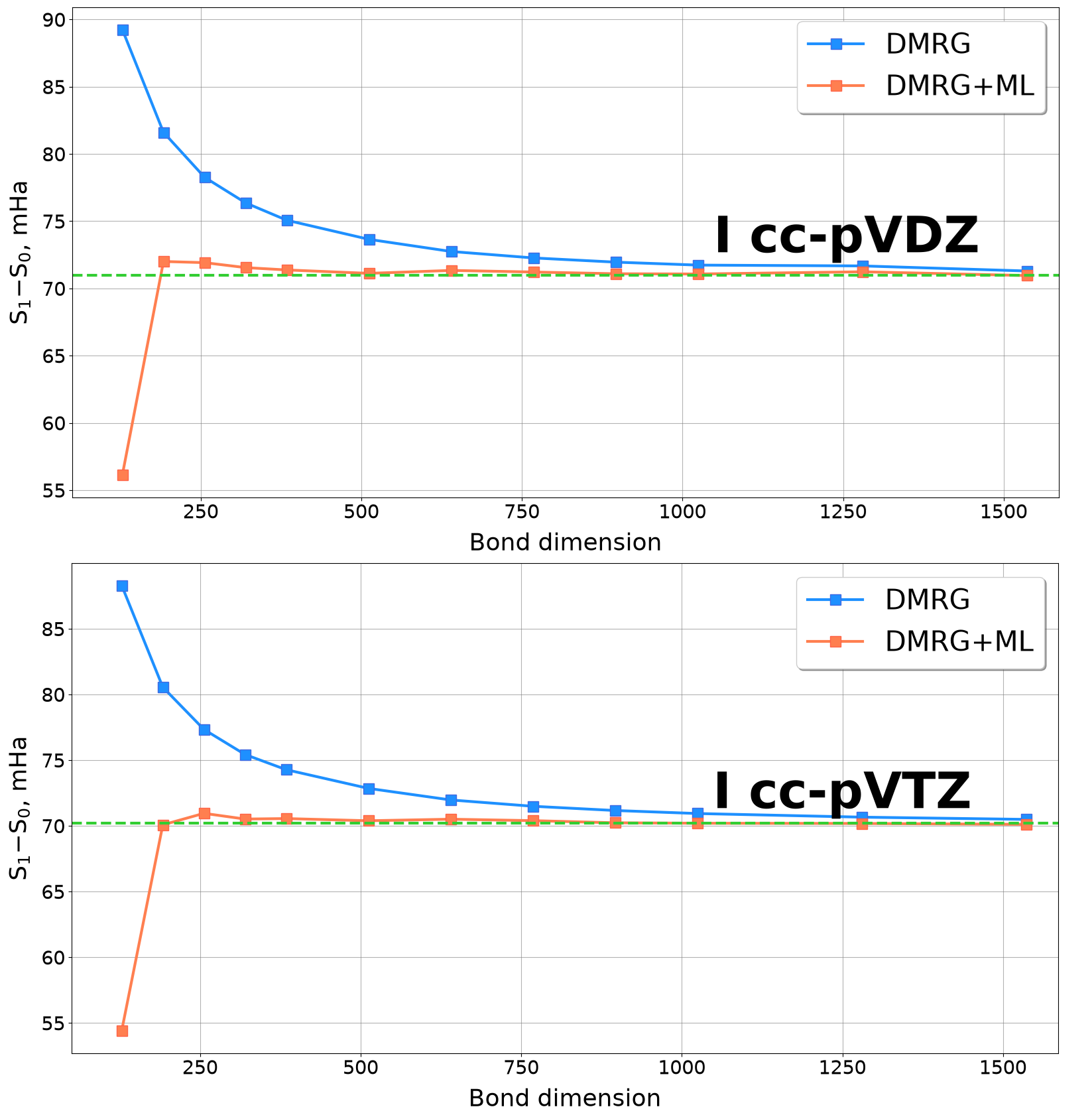}
  \caption{$S_1 - S_0$ energy gap for octacene, C\textsubscript{34}H\textsubscript{20}. The reference energy gap (green line) is obtained with $D=3072$, 71.0 mHa for cc-pVDZ, 70.2 mHa for cc-pVTZ.}
  \label{C34H20_pred}
\end{figure}

\begin{figure}[!t]
  \includegraphics[width=8cm]{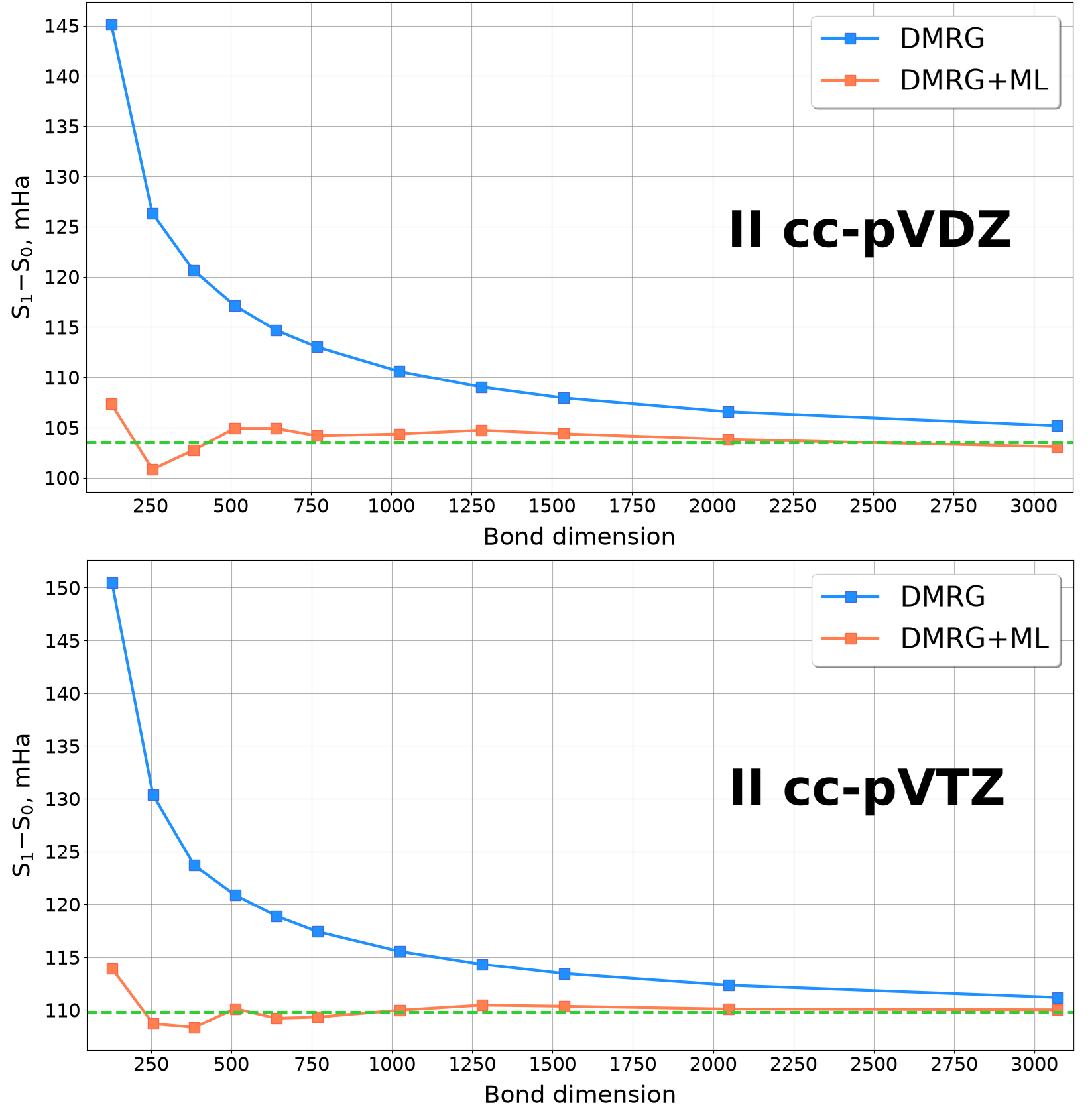}
  \caption{$S_1 - S_0$ energy gap for ovalene, C\textsubscript{32}H\textsubscript{14}. The reference energy gap (green line) is obtained with $D=8192$, 103.5 mHa for cc-pVDZ, 109.8 mHa for cc-pVTZ.}
  \label{C32H14_pred}
\end{figure}

\begin{figure}[!t]
  \includegraphics[width=8cm]{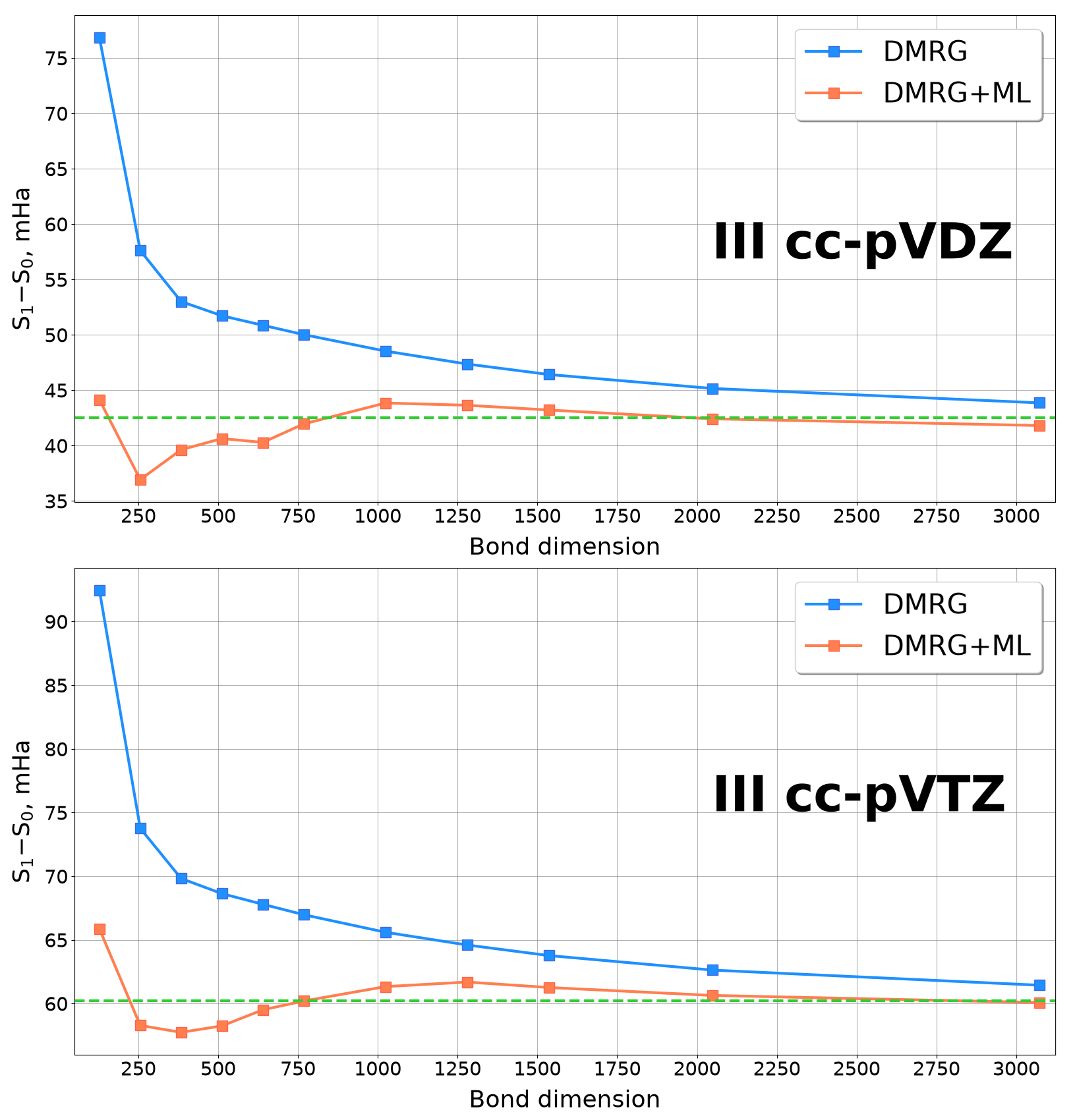}
  \caption{$S_1 - S_0$ energy gap for aza-analogue of [4]triangulene, C\textsubscript{26}N\textsubscript{7}H\textsubscript{11}. The reference energy gap (green line) is obtained with $D=8192$, 42.5 mHa for cc-pVDZ, 60.2 mHa for cc-pVTZ.}
  \label{C26N7H11_pred}
\end{figure}

\begin{figure}[!t]
  \includegraphics[width=8cm]{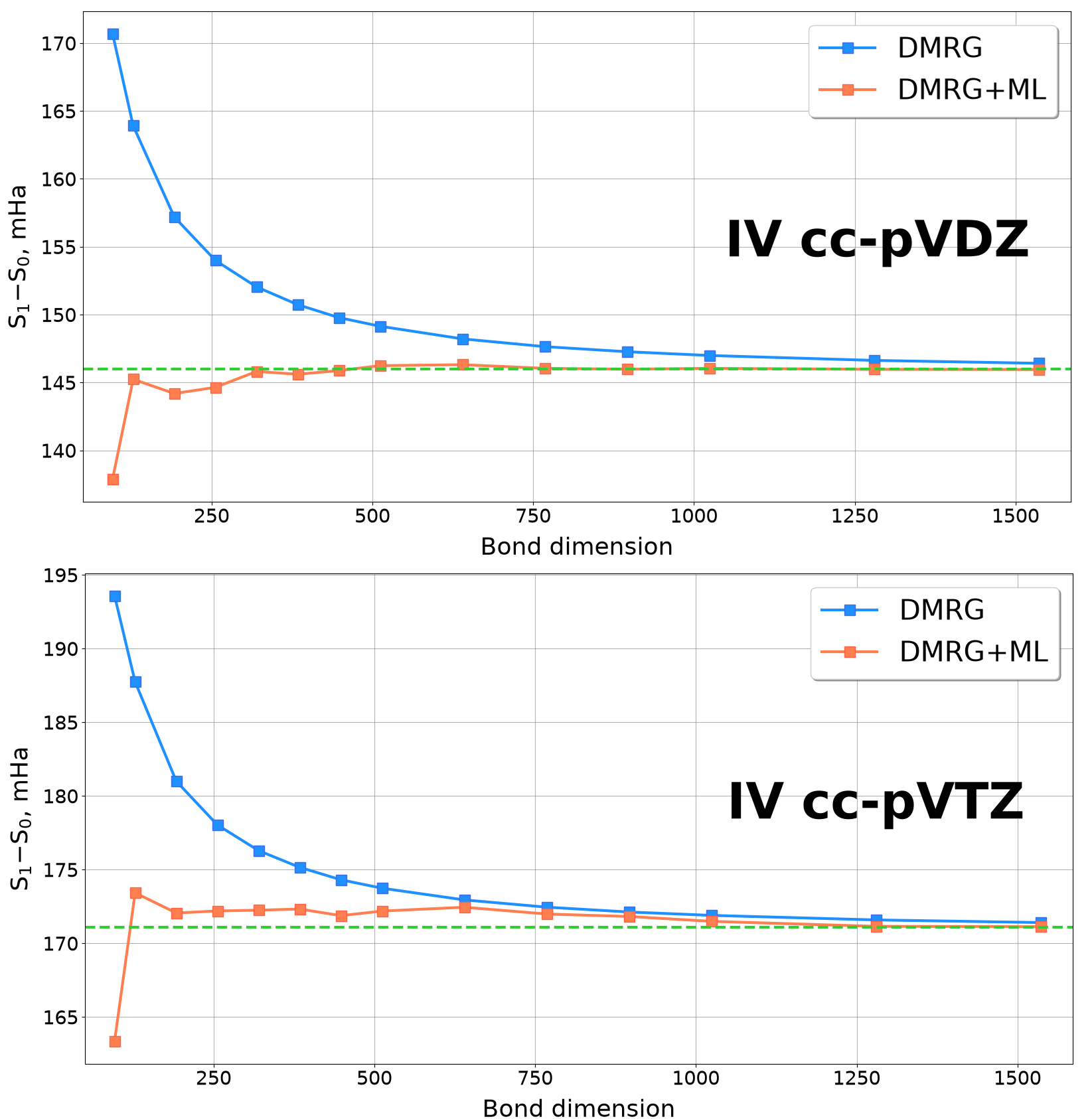}
  \caption{$S_1 - S_0$ energy gap for 4,11-diazadibenzo[\textit{g,p}]chrysene, C\textsubscript{24}N\textsubscript{2}H\textsubscript{14}. The reference energy gap (green line) is obtained with $D=3072$, 146.0 mHa for cc-pVDZ, 171.1 mHa for cc-pVTZ.}
  \label{C24N2H14_pred}
\end{figure}

\begin{figure}[!t]
  \includegraphics[width=8cm]{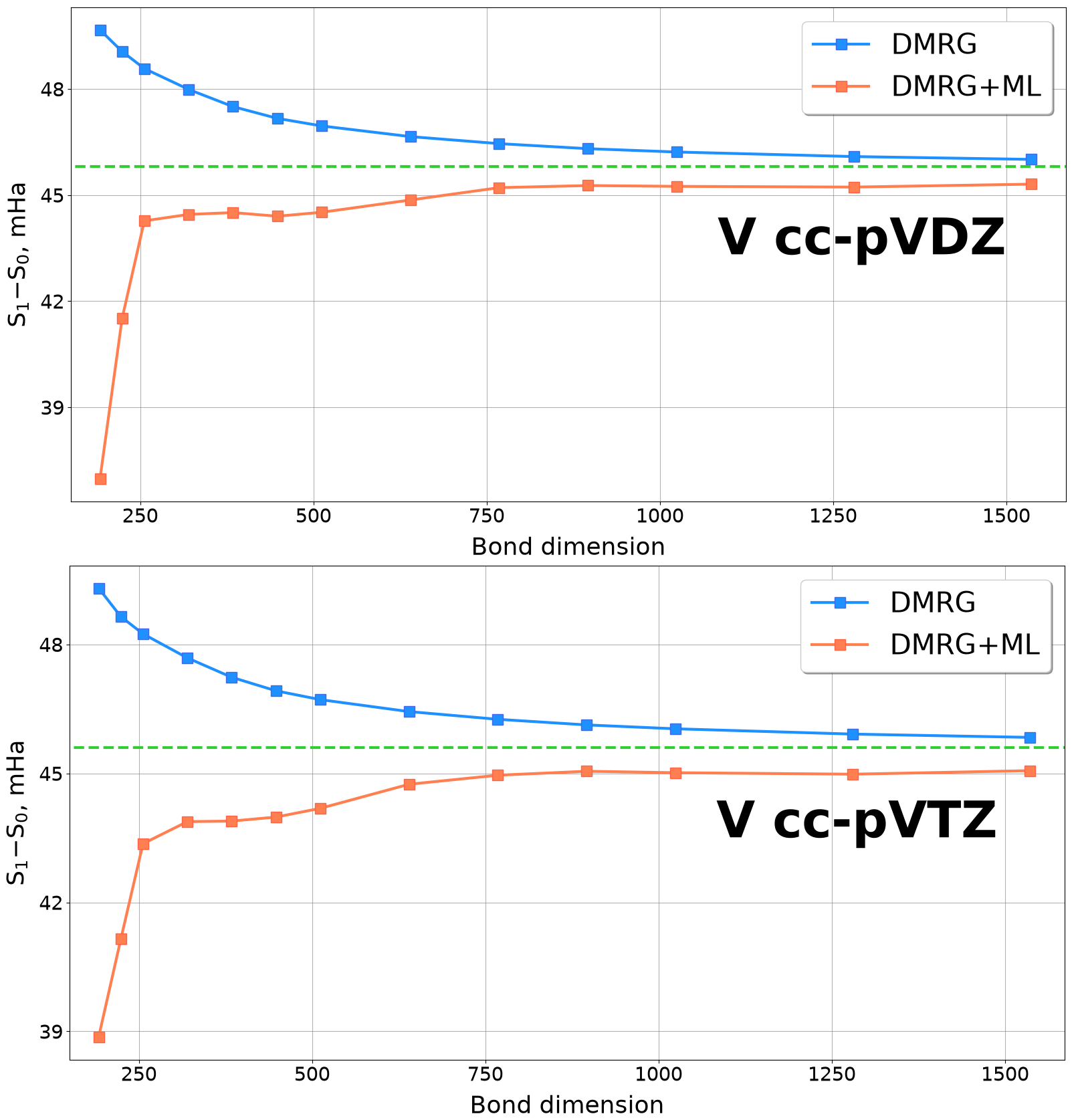}
  \caption{$S_1 - S_0$ energy gap for bis(phenalenyl) linked by benzene, C\textsubscript{32}H\textsubscript{16}. The reference energy gap (green line) is obtained with $D=8192$, 45.8 mHa for cc-pVDZ, 45.6 mHa for cc-pVTZ.}
  \label{C32H16_pred}
\end{figure}

\subsection{ML model performance}

The best performance of the ML model is detected for octacene (Fig. \ref{C34H20_pred}). The ML refinement accuracy within 1 mHa is shown even for small bond dimensions, except for the very small, $D < 200$. This is expected since the general structure principle of the systems in the training dataset can be imagined as various combinations of acene and phenacene motifs of different sizes. In the same time, the excellent performance on the acene which is larger than any example from the training set, proves that the model correctly captures the electronic structure/size dependence for acenes and phenacenes.

The performance of the model for ovalene  (Fig. \ref{C32H14_pred}), aza-analogue of [4]triangulene  (Fig. \ref{C26N7H11_pred}), and 4,11-diazadibenzo[\textit{g,p}]chrysene (Fig. \ref{C24N2H14_pred}) is similarly good and comparable across the three systems. Although, at small $D$, the error in the energy refinement can reach up to 6 mHa, it generally decreases to approximately 1.6 mHa ($\approx 1$ kcal/mol) at larger $D$, corresponding to a TRE in the range of $3.0$--$1.0\times10^{-4}$. Depending on the system, this TRE range corresponds to bond dimensions between approximately 200 and 500.

The most challenging example is proved to be bis(phenalenyl) (Fig. \ref{C32H16_pred}). At some very small $D$ the prediction can even be worse than the actual DMRG result. It should be underlined that the correlation pattern of ground singlet state in this case is much different from the correlation patterns of the previous examples and compounds in the learning dataset. Indeed, the latter are characterized by a relatively even distribution, i.e. all split-localized $\pi$-orbitals have $s^{(1)}$ values close to each other -- in the range $0.2-0.3$ or $0.4-0.5$. In contrast, in bis(phenalenyl) $\pi$- and $\pi^{*}$-orbitals localized on the connector and adjacent parts of phenalenyl radicals are characterized by very high values of $s^{(1)}$ -- in the range $0.6-1.0$, while $\pi$-orbitals localized on distant parts of phenalenyl radicals have $s^{(1)}$ close to $0.2$.

However, even for bis(phenalenyl) the ML model starts to improve $S_0 \rightarrow S_1$ gaps at bond dimensions from 250 to 400. This corresponds to the TRE in the range $2.2 - 1.2 \times 10^{-4}$, which normally for the case of ML unrefined DMRG calculations are deemed as an inaccurate outcome. Further, all predictions continue to stay within 1 kcal/mol, that is, within the general accuracy of the model.

In general, the ML model allows to improve the $S_0 \rightarrow S_1$ transition energies to within 1 mHa starting with moderately small values of $D$ and large values of TRE $2 \times 10^{-4} - 7 \times 10^{-5}$. This offers significant savings in computational costs, since comparable accuracy without ML refinement requires a set up of 3000 -- 4000 $D$ for systems I and IV, and even 7000 -- 8000 $D$ for systems II, III, and V. In the same time, seemingly it is unlikely to obtain systematically such performance on extremely small bond dimensions, since for strongly-correlated systems an important part of correlation coupling might be truncated as well, rendering faulty numerical results and as a consequence an inconsistent input for the ML model.

The data are also available in the form of tables in Supporting materials.

\section{Conclusions}

We present a machine learning scheme for improving the energies of the ground and first excited states of the same spin multiplicity, described within the CAS ansatz using DMRG as the active-space solver. The resulting ML model is trained on complete $\pi$-valence active spaces of aromatic hydrocarbons and tested to predict the corresponding $S_0 \rightarrow S_1$ transition energies. The model provides an improvement of low- and medium-quality DMRG calculations to a high-accuracy level with error within 1 kcal/mol, which tends systematically to decrease further with increasing of the quality of DMRG calculations (i.e. with increasing of bond dimension). The model shows good performance on a wide range of hydrocarbons featuring different electronic structures: planar laterally and peri-fused systems, hetero- and chiral hydrocarbons, or diradicals.

Although the model is applicable \textit{per se} to polycyclic aromatic hydrocarbons, the principles of its design matter no less, if not more. Indeed, the model design and the constituent components are not dependent on the specific compound class or excitation type, and the ML workflow (Figs. \ref{gnn_scheme}, \ref{learning_scheme}) is invariant with respect to the size or type of system. It can be straightforwardly generalized and re-parametrized to multiple states, non-vertical excitations, or other types of strongly correlated systems. However, with the latter, the problem of finding a procedure for reliable separation and localization of strongly-correlated components from generally delocalized canonical wave function should be addressed.

The same quality performance of the ML model on both double and triple zeta basis sets makes it possible to use it as a component part of different approaches for complete basis set extrapolation. The model is available online through the GitHub repository.\cite{github}

\section{Computational and learning details}

The training data set has been composed of elements from the publicly available database of polycyclic aromatic hydrocarbons COMPAS-1D\cite{Wahab-Pfuderer-2022}. 100 molecules have been selected: all hydrocarbons with 5 and 6 benzene rings (49 molecules in total) and 51 hydrocarbons with 7 benzene rings and the smallest HOMO -- LUMO gaps.

The input orbitals for the DMRG calculations have been computed at the DFT level using the B3LYP exchange-correlation functional \cite{Parr88_785,Becke88_3098} and the cc-PVDZ basis set \cite{Dunning1989}. The orbitals have been split-localized (separate localization
of occupied and virtual orbitals) with the Pipek-Mezey method \cite{Pipek1989}. Complete $\pi$-active spaces (22e, 22o), (26e, 26o), and (30e, 30o) have been selected for molecules with 5, 6, and 7 benzene rings (5C, 6C, and 7C) respectively. The ordering of the active orbitals mapped onto the 1D lattice of the DMRG sites has been obtained with the Fiedler method \cite{Barcza2011} applied to the matrix of exchange integrals \cite{Olivares-Amaya-Hu-2015}. All DMRG calculations have been initialized with the CI-DEAS procedure \cite{Legeza-Solyom-2003} and performed with MOLMPS program \cite{Brabec2020, molmps_scalable}.

DMRG calculations with bond dimensions 128, 192, 256, 384, 512, 640, 768, 896, 1024 for all molecules and with bond dimension 2048 for 6C and 7C molecules have been used as training points. This ensured the input data with the truncation error in the 5$\times$10$^{-4}$ -- 1$\times$10$^{-5}$ range. The reference targets have been obtained with bond dimension 2048 for 5C molecules and with bond dimension 3072 for 6C and 7C molecules.

The target energy values have been taken as the energy difference between the Hartree-Fock energy of the input orbitals and the corresponding DMRG optimized energy.

The multi-layer perceptron unit at stage \textit{MP1} (Fig. \ref{learning_scheme}) was fully-connected 3-layer neural network with 100 neurons in each layer, featuring sigmoid linear unit (SiLU) activation ($f(z) = z (1 + e^{-z})^{-1}$) for each neuron. The multi-layer perceptron unit at stage \textit{MP2} (Fig. \ref{learning_scheme}) was a single-layer fully-connected neural network with 40 neurons featuring SiLU activation function for each neuron. An output neuron features leaky ReLU ($f(z) = max \{ 0,z \} + 0.2 \cdot min \{ 0,z \}$) activation function. To avoid overfitting, the dropout regularization procedure\cite{Hinton-Srivastava-at-al-2012, Srivastava-Hinton-at-al-2014} has been used with a dropout rate of 0.1. Each train procedure run over 2000 epochs.

Test molecules have been subject to the same active space selection procedure, which resulted in the following $\pi$-active space configurations: {\bf{I}} -- C$_{34}$H$_{20}$, (34e, 34o); {\bf{II}} -- C$_{32}$H$_{14}$, (32e, 32o); {\bf{III}} -- C$_{26}$N$_{7}$H$_{11}$, (36e, 33o); {\bf{IV}} -- C$_{24}$N$_{2}$H$_{14}$, (26e, 26o); {\bf{V}} -- C$_{32}$H$_{16}$, (32e, 32o). Their geometries have been optimized at the DFT level using the B3LYP exchange-correlation functional and the cc-PVTZ basis set and are available in the Supporting materials.

\section{Acknowledgement}
This work was supported by the Czech Science Foundation (grant no.\ 25-18486S) and
the Ministry of Education, Youth and Sports of the Czech Republic through the e-INFRA CZ (ID:90254).

\bibliography{references} 

\end{document}